\documentclass[a4paper]{jpconf}
\usepackage{graphicx}
\begin{document}
\title{Geodynamo and mantle convection simulations on the Earth Simulator using the Yin-Yang grid}

\author{Akira Kageyama and Masaki Yoshida}

\address{The Earth Simulator Center,
Japan Agency for Marine-Earth Science and Technology,
Showa-machi 3173-25, Yokohama, Japan}

\ead{kage@jamstec.go.jp}

\begin{abstract}
We have developed finite difference codes based on the Yin-Yang grid for 
the geodynamo simulation and the mantle convection simulation.
The Yin-Yang grid is a kind of spherical overset grid that is composed of
two identical component grids.
The intrinsic simplicity of the mesh configuration of the Yin-Yang grid enables us to develop highly optimized simulation codes on massively parallel supercomputers.
The Yin-Yang geodynamo code has achieved 15.2 Tflops with 4096 processors on 
the Earth Simulator.
This represents 46\% of the theoretical peak performance.
The Yin-Yang mantle code has enabled us to carry out mantle
convection simulations in realistic regimes 
with a Rayleigh number of $10^7$ including
strongly temperature-dependent viscosity with spatial contrast up to $10^6$.
\end{abstract}

\section{Introduction}
%
The Earth (radius $r=6400$km)
is composed of three spherical layers;
the inner core ($r=1200$km), 
the outer core ($r=3500$km),
and the mantle.
Computer simulations of the Earth's interior need efficient spatial 
discretization methods in the spherical shell geometry.
To achieve high sustained performance on massively parallel supercomputer
such as the Earth Simulator,
spatially localized discretization methods rather than spectral methods
are desirable. 
Recently, we proposed a new spherical grid system, the ``Yin-Yang grid,''
for geophysical simulations.
Because there is no grid mesh that is orthogonal 
over the entire spherical surface and, 
at the same time, free of coordinate singularity or grid convergence, 
we have chosen an overset grid approach.
A spherical surface is decomposed into two identical subregions. 
The decomposition (or dissection) enables us to cover each 
subregion by a grid system that is individually orthogonal and singularity-free.
Each component grid in this Yin-Yang grid is a low latitude 
component of the usual latitude-longitude grid on the spherical 
polar coordinates (90 degree about the equator and 270 degree in the longitude).
Therefore, the grid spacing is quasi-uniform and the metric tensors 
are simple and analytically known.
One can directly apply mathematical and numerical 
resources that have been written in the 
spherical polar coordinates or latitude-longitude grid.
Since the two component grids are identical 
and they are combined in a complementary way, 
various routines of the code can be recycled twice 
for each component grid at every simulation time step.
We have developed finite difference codes based on 
the Yin-Yang grid for (i) the geodynamo simulation in the outer core, 
and (ii) the mantle convection simulation.

In general, a dissection of a computational domain generates
internal borders or internal boundaries  between the subregions.
In the overset grid methodology \cite{chesshire_1990},
the subregions are permitted to partially overlap one another on their borders.
The overset grid is also called as 
overlaid grid,
or composite overlapping grid,
or Chimera grid \cite{steger_1983}.
The validity and importance of the overset approach in
the aerodynamical calculations was
pointed out by Steger \cite{steger_1982}.
Since then
this method is widely used in this field.
It is now one of the most important grid techniques
in the computational aerodynamics.

In the computational geosciences,
the idea of the overset grid approach appeared 
rather early.
Phillips proposed a kind of composite grid
in 1950's to solve partial differential equations
on a hemisphere,
in which the high latitude region of 
the latitude-longitude grid is
``capped'' by another 
grid system that is constructed by a 
stereographic projection to a plane 
on the north pole \cite{phillips_1957,phillips_1959,browning_1989}.
After a long intermission,
the overset grid method seems to attract growing interest
in geoscience these days.
The ``cubed sphere'' \cite{ronchi_1996} 
is an overset grid that covers a spherical 
surface with six component grids that
correspond to six faces of a cube.
The ``cubed sphere'' is recently applied to the 
mantle convection simulation \cite{hernlund_2003}.
In the atmospheric research,
other kind of spherical overset grid is used
in a global circulation model \cite{dudhia_2002},
in which the spherical surface is covered
by two component 
grids---improved stereographic projection grids---in
northern and souther hemispheres that
overlap in the equator.

Among indefinite variations of spherical overset grid systems,
what is the simplest one?
In general, 
the structure of a spherical overset grid is largely determined by 
the number of divided pieces of the sphere $n$ ($\ge 2$).
Here we consider the minimum case of $n=2$, i.e., 
the spherical dissections by two pieces.
One can divide a sphere into two parts, for example, by cutting
along a small circle at any latitude.
We concentrate on a special class of $n=2$ dissections
in which the two pieces are geometrically identical, i.e.,
they have exactly same size and shape.
Another condition we impose here to maximize the simplicity
is the symmetry of the piece.
It should have two fold symmetry in two perpendicular directions;
up-down and right-left.
Here we call this special class of dissections as yin-yang dissection of a sphere.

A trivial example of the yin-yang dissection is
obtained by cutting along the equator or any great circle,
producing two hemispheres.

\begin{figure}[ht]
\includegraphics[width=18pc]{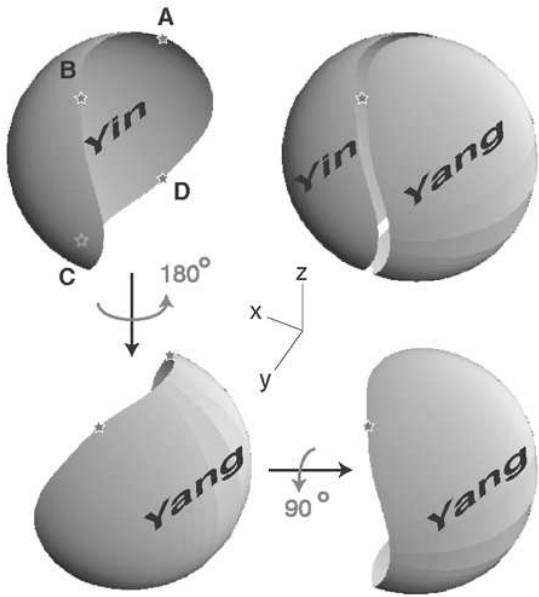}\hspace{2pc}%
\begin{minipage}[b]{18pc}
\caption{\label{fig:sphericalDissection}
An example of yin-yang dissection of a sphere:
A sphere is divided into two identical pieces,
with same shape and size.
Each piece has two fold symmetry; up-down and right-left.
They are combined in a complemental way to cover a spherical surface.
The two identical pieces of the yin-yang dissection is transformed
each other by two successive rotations, or one rotation.
}
\end{minipage}
\end{figure}

Other yin-yang dissections are obtained by modifying the cut curve from the great circle.
Let $S_{yin}$ be a piece of a sphere $S$ with radius $r=\sqrt{2}$.
We should keep the surface area of $S_{yin}$ being
$2\pi r^2$, just a half of $S$'s surface.
An example of $S_{yin}$ is shown in the upper left panel in
Fig.~\ref{fig:sphericalDissection}.
The border curve of $S_{yin}$
passes through the following four points on the sphere;
point $A$ at $(x,y,z)=(0,-1,+1)$,
$B$ at $(0,+1,+1)$,
$C$ at $(0,+1,-1)$
and $D$ at $(0,-1,-1)$.
The curve $AB$, between $A$ and $B$, 
is arbitrarily as long as it
is symmetric about the $y=0$ plane.
Other three curves, $BC$, $CD$, and $DA$, are uniquely constructed from 
the curve $AB$ as follows:
The curve $BC$ is a copy of $AB$ followed by two successive rotations,
first $180$ degree about the $z$~axis, 
then $90$ degree about the $x$~axis.
The curve $CD$ is the mirror image of $AB$ about $z=0$ plane.
The curve $DA$ is the mirror image of $BC$ about $y=0$ plane.
From this definition of the border curve $ABCD$, it is obvious that the surface
area of $S_{yin}$ is just a half of that of the sphere $S$.
Now we make a copy of $S_{yin}$ and call it $S_{yang}$
which is rotated for $180^\circ$ around $z$-axis.
(See lower left panel of Fig.~\ref{fig:sphericalDissection}.)
Then, rotate it again, but this time for $90^\circ$ degree around $x$-axis,
as shown in the lower right panel.
Then the original piece $S_{yin}$ (the upper left) and 
the rotated copy $S_{yang}$  (the lower right)
can be combined, and they just cover the original sphere $S$ as
shown in the upper right in this figure.
This is an constructive illustration of the yin-yang dissection of a sphere.

Since the initial curve $AB$ was arbitrarily,
it is obvious that
there are indefinite variations of the yin-yang dissection of the sphere $S$.

\section{Yin-Yang grids}

\begin{figure}[ht]
\includegraphics[width=18pc]{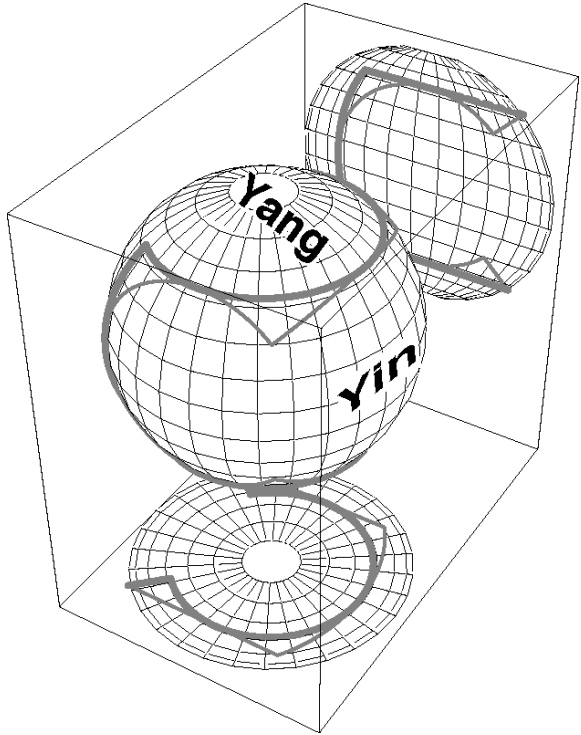}\hspace{2pc}%
\begin{minipage}[b]{18pc}
\caption{\label{fig:curveBasicYinYang}
A dissection of a sphere into two identical pieces---Yin and Yang---with 
a partial overlap.
The thick and thin curves are the borders of Yin and Yang piece, 
respectively.
the thick curve (Yin's border) is always located in either constant latitudes or constant longitudes.
The Yin (Yang) piece is a rectangle in the computational ($\theta,\phi$) space of the Yin (Yang) grid.
}
\end{minipage}
\end{figure}

\begin{figure}[ht]
\begin{center}
\noindent\includegraphics[width=\linewidth]{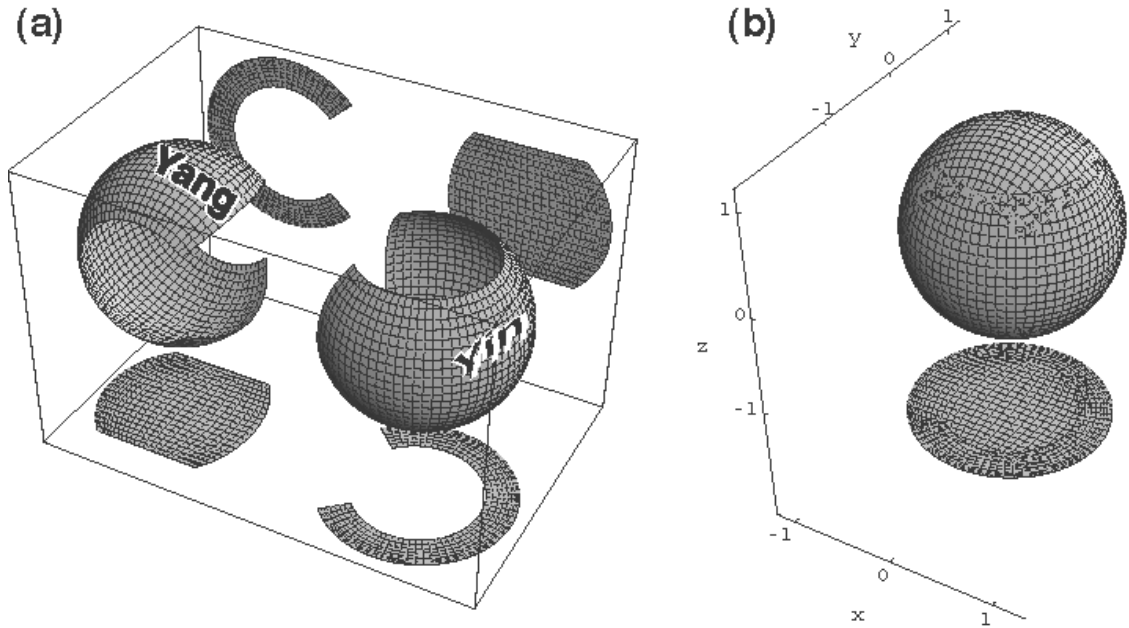}
\caption{
\label{fig:yinyangBasic}
A Yin-Yang grid based on 
 the yin-yang dissection with partial overlap shown in Fig.~\ref{fig:curveBasicYinYang}.
Each component grid is rectangle in the computational ($\theta,\phi$) space.
}
\end{center}
\end{figure}

The overset grid methodology gives us a freedom
to design the shape of the component grid as long as
the grids has \textit{minimum} overlap one another \cite{chesshire_1990}.
Therefore, we can take the component grid as
a rectangle in the computational ($\theta, \phi$) space.
Fig.~\ref{fig:curveBasicYinYang} shows a spherical 
dissection by two identical pieces with partial overlap.
The Yin piece is surrounded by a thick red curve and Yang piece is
surrounded by a thin curve.
Note that the northern and southern borders of the Yin piece
are located in constant latitudes and
the western and eastern borders are located in constant longitudes.
In other words,
the Yin piece in Fig.~\ref{fig:curveBasicYinYang} 
is a rectangle in the ($\theta, \phi$) space
of the Yin's spherical coordinates, and therefore, 
the Yang piece is also (the same)
rectangle in Yang's coordinates that is perpendicular to the Yin's.
The Yin-Yang grid based on this partially overlapped spherical dissection
is shown in Fig.~\ref{fig:yinyangBasic}.
Here, each component grid spans 
the subregion $S_y$  defined by
\begin{equation} \label{eq:a00}
  S_y := \{\theta,\phi\},\ 
                  |\, \theta-\pi/2\, | \le \pi/4 + \delta, \, 
                  |\, \phi \, | \le 3\pi/4 + \delta,
\end{equation}
with a small buffer $\delta$ which is necessary to keep the minimum
overlap between Yin and Yang.
Note that in the simulation code, 
one subroutine for the fluid solver, for instance, can be
recycled twice because the grid distribution is exactly the
same for the Yin and Yang.

The Yin and Yang are
converted each other by a rotation.
The Yin's cartesian coordinates $x^n_i$ for $i=1,2,3$ and
that Yang's coordinates $x^e_i$ are related by
\begin{equation} \label{eq:a05}
   x^e_i = M_{ij} x^n_j \ \ \ \hbox{for}\ i, j = 1,2,3
\end{equation}
where $M_{11} = -1$, $M_{23}=M_{32}=1$,
and $M_{ij}=0$ for other components.
Note that the matrix $M$ satisfies 
\begin{equation}
  M = M^t = M^{-1},                          \label{eq:a06}
\end{equation}
which indicates a complemental relation between
the Yin and Yang.
The coordinate transformation from Yin to Yang is
mathematically the same as that from Yang to Yin.
This enables us
to make only one, instead of two, subroutines that involve  any 
data transformation between Yin and Yang,
which is required in the mutual interpolation for the internal
boundary condition on the overset grid borders.

The transformation formula of any vector components
$\mathbf{v}=(v_r, v_\theta, v_\phi)$ between Yin and Yang is given by
\begin{equation}
  \left( 
    \begin{array}{c}
      v_r^e \\
      v_\theta^e\\
      v_\phi^e
    \end{array}
  \right) 
  =
  P
  \left(
    \begin{array}{c}
      v_r^n \\
      v_\theta^n \\
      v_\phi^n
    \end{array}
  \right),
\end{equation}
with the transformation matrix
\begin{equation}
  P =
  \left(
    \begin{array}{ccc}
      1 & 0 & 0 \\
      0 & -\sin{\phi^e} \sin{\phi^n} & - \cos{\phi^n} / \sin{\theta^e}\\
      0 & \cos{\phi^n} / \sin{\theta^e} & -\sin{\phi^e} \sin{\phi^n} 
    \end{array}
  \right).
\end{equation}
The inverse transformation is given by the same matrix; $P^{-1}=P$, 
which is another reflection of the complemental nature between the Yin and Yang.

Another merit of the Yin-Yang grid
resides in the fact that the component grid is 
nothing but (a part of ) the latitude-longitude grid.
We can directly deal with the equations to be solved with 
the usual spherical polar coordinates.
The analytical form of metric tensors are familiar
in the spherical coordinates.
We can directly code
the basic equations 
in the program as they are formulated in the
spherical coordinates.
We can make use of various resources of
mathematical formulas, program libraries, and other tools 
that have been developed in the spherical polar coordinates.

In order to illustrate the programing strategy in the Yin-Yang method,
let us consider a two-dimensional fluid problem on a sphere $S$.
Suppose that two components of the flow velocity $\mathbf{v}=(v_\theta, v_\phi)$
and the pressure $p$ 
are written in \texttt{vel\_t}, \texttt{vel\_p}, and \texttt{press} in a Fortran~90/95 program.
They can be combined into one structure or ``type'' in Fortran~90/95 as
\begin{quote}
{\ttfamily
type fluid\_ \\
 \ \  real(DP), dimension(NT,NP) :: vel\_t, vel\_p, press  \\
end type fluid\_ 
}     
\end{quote}
where \texttt{NT}, \texttt{NP} are the grid size integers
in $\theta$ and $\phi$ directions in 
the subregion $S_y$ of eq.~(\ref{eq:a00}).
Using this structured type,
we declare two variables for the fluid; one 
is for Yin and another for Yang:
\begin{quote}
{\ttfamily
type(fluid\_) :: fluid\_yin, fluid\_yang
}     
\end{quote}
Then, we call a fluid solver subroutine, here named \texttt{navier\_stokes\_solver},
that numerically solves the Navier-Stokes equation 
in the spherical coordinates in the subregion $S_y$:
\begin{quote}
{\ttfamily
call navier\_stokes\_solver(fluid\_yin)\\
call navier\_stokes\_solver(fluid\_yang)
}     
\end{quote}
The first call of \texttt{navier\_stokes\_solver}
solves the fluid motion in the $S_y$ region defined in the Yin's spherical coordinates
and the second call is for the same region $S_y$ defined
in the Yang's coordinates.
But in the program code, we do not have to distinguish
the two $S_y$ regions since the basic equations,
numerical grid distribution, and therefore, all numerical tasks
are identical in the computational space.
For a rotating fluid problem with a constant angular velocity $\mathbf{\Omega}$,
we have the Coriolis force term in the Navier-Stokes equation
that seems to break the symmetry between the Yin grid and Yang grid,
but it is still possible to write the equation in exactly the same form
for the Yin and Yang grids by explicitly
writing three components of angular velocity in the
Coriolis force term  $2 \mathbf{v}\times \mathbf{\Omega}$ in the subroutine.
Then, we call the routine with the angular velocity vector
in each grid (Yin or Yang)
as the second argument:
\begin{quote}
{\ttfamily
call navier\_stokes\_solver(fluid\_yin,omega\_yin)\\
call navier\_stokes\_solver(fluid\_yang,omega\_yang)
}     
\end{quote}
where \texttt{omega\_yin} and \texttt{omega\_yang}
are again structured variables that hold three components
of the $\mathbf{\Omega}$ vector:
For example,
 \texttt{omega\_yin} holds three components of cartesian vector components
in the Yin grid
$(\Omega_x^n, \Omega_y^n, \Omega_z^n)=(0,0,\Omega)$,
and
\texttt{omega\_yang} holds
$(\Omega_x^e, \Omega_y^e, \Omega_z^e)=(0,\Omega,0)$.

Our experience tells that
it is easy to convert an existing 
latitude-longitude based program into a Yin-Yang based program since
there are many shared routines between them.
In addition to that the size of the code as well as its complexity
is drastically reduced by the code conversion because
we can remove routines that are designed to resolve the pole problems
on the latitude-longitude grid.

%
\section{Application to the mantle convection simulation}
%

\subsection{Simulation model}
We applied the Yin-Yang grid described in the previous secion
for the mantle convection simulation.
The details of the adopted numerical methods 
and benchmark tests can be found in \cite{yoshida_2004}.

We model the mantle convection as a thermal convection 
of a Boussinesq fluid with infinite Prandtl number 
heated from bottom of a spherical shell~\cite{mckenzie_1974}. 
The ratio of the inner radius ($r=r_0$)
and the outer radius ($r=r_1$) is 0.55. 
The normalization factors for the non-dimensionalization of 
the length, velocity, time 
and temperature are $\hat{d}$, $\hat{\kappa}/\hat{d}$, 
$\hat{d}^2 / \hat{\kappa}$ and 
$\Delta \hat{T} = \hat{T}_{bot} - \hat{T}_{top}$, respectively,  
where 
$d$ is the thickness of the shell, 
$\hat{\kappa}$ the thermal diffusivity, 
and $\hat{T}_{bot}$ and $\hat{T}_{top}$ are the temperatures 
on the bottom and top surfaces. 
The hat stands for dimensional quantity. 
The non-dimensional equations of mass, momentum, and energy conservation 
governing the thermal convection are,  
\begin{equation}
\mathbf{\nabla} \cdot \mathbf{v} = 0,
\end{equation}
\begin{equation}
  0 = - \mathbf{\nabla} p + \mathbf{\nabla} \cdot ( \mu \mathbf{\dot{e}} ) 
    + Ra T \hat{\mathbf{r}}, 
\end{equation}
\begin{equation}
  \frac{\partial T}{\partial t} = \mathbf{\nabla}^2 T  
  - \mathbf{v} \cdot \mathbf{\nabla} T + H,
\end{equation}
where $\mathbf{v}$ is the velocity vector,  
$p$ pressure, 
$\mu$ viscosity,
$T$ temperature, $t$ time, 
$\mathbf{\dot{e}}$ strain-rate tensor, 
and $\hat{\mathbf{r}}$ is the unit vector in the $r$-direction. 
The Rayleigh number is defined by
\begin{equation}   
  Ra \equiv \frac{\hat{\rho} \hat{g} \hat{\alpha} \Delta \hat{T} \hat{d}^3}
                 {\hat{\kappa} \hat{\mu}},
\end{equation}
where  
$\hat{\rho}$ is the density, 
$\hat{g}$ the gravitational acceleration, and 
$\hat{\alpha}$ is the thermal expansivity. 
Most of the heat for Earth's mantle comes 
from a combination of radioactive decay of isotopes 
and secular cooling of the mantle. 
The internal heating is defined by
\begin{equation}      
  H \equiv \frac{\hat{Q} \hat{d}^2}{\hat{\kappa} \hat{c}_p \Delta \hat{T}},
\end{equation} 
where $\hat{Q}$ is the internal heating rate per unit mass, and 
$\hat{c}_p$ is the specific heat at constant pressure.  

According to the laboratory experiments on silicate rock deformation, 
the viscosity of the Earth's mantle depends on various parameters 
such as temperature, pressure, stress, and so on~\cite{turcotte_2002}. 
Among them, temperature dependence is the most dominant factor. 
Here we assume that viscosity $\mu$ depends only on temperature;  
\begin{equation}
  \mu(T) = \exp \left[ -E \left( T - T_{bot} \right) \right].
\end{equation}
The viscosity contrast across the spherical shell is defined by 
$\gamma_{\mu} \equiv \mu(T_{top}) / \mu(T_{bot}) = \exp(E)$. 
The mechanical boundary conditions at the top  
and bottom surface are immpermiable and stress-free. 
The boundary conditions for $T$ are fixed; 
$T_{bot} = 1$ and $T_{top} = 0$.

\begin{figure}[ht]
\begin{center}
\noindent\includegraphics[width=0.8\linewidth]{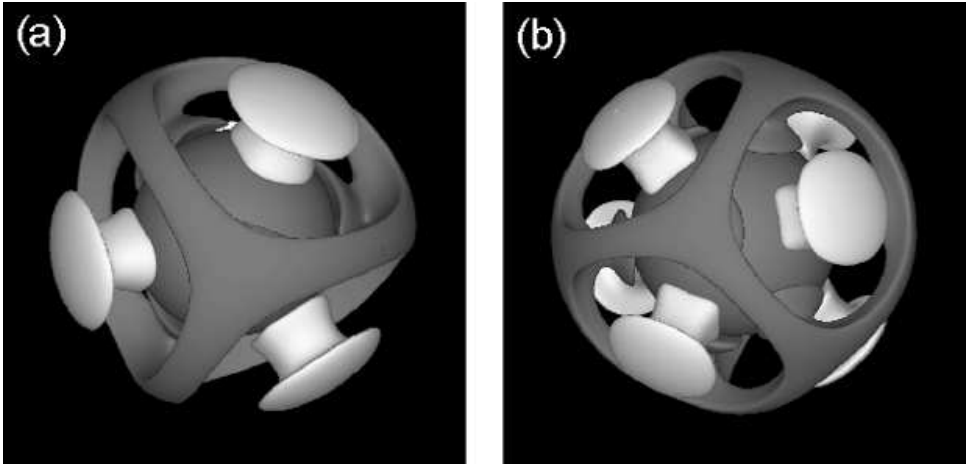}
\caption{
\label{fig:yoshida01}
The iso-surface of the residual temperature $\delta T$
(the deviation from horizontally averaged temperature at each depth) 
started from the initial condition of 
(a) the tetrahedral and (b) the cubic symmetries. 
The Rayleigh number is $Ra = 10^4$. 
Blue and Yellow iso-surfaces indicate $\delta T = -0.125$ and $\delta T = 0.150$, respectively.  
Red spheres indicate the bottom of the mantle with fixed temperature.
}
\end{center}
\end{figure}

\subsection{Steady state convection}

The thermal convection in the spherical shell with infinite Prandtl number 
has two stable solutions with polyhedral symmetry when the Rayleigh number is low.   
The two solutions are found by linear theory and confirmed 
by numerical simulations~\cite{bercovici_1989}: 
One solution is a convection with the tetrahedral symmetry which has four upwellings;   
the other has the cubic symmetry with six upwellings. 
To confirm these symmetric solutions and their stabilities,  
we performed two simulations with different initial conditions of temperature field;   
$T(r, \theta, \phi) = T_{cond}(r) + T_{prtb}(r, \theta, \phi)$, 
where $T_{cond}(r)$ is the purely conductive profile, $\mathbf{\nabla}^2 T_{cond}(r) = 0$,  
with the thermal boundary conditions given above. 
The perturbation term $T_{prtb}(r, \theta, \phi)$ is given by,  
\begin{eqnarray}
  T_{prtb}(r, \theta, \phi) =  0.1\, Y_3^{2}(\theta, \phi) \sin \pi (r - r_0),
\end{eqnarray}
for the tetrahedral symmetric solution, and 
\begin{eqnarray}
  T_{prtb}(r, \theta, \phi) =  0.1 \left\{  {Y_4}^{0} (\theta, \phi) 
                            + \frac{5}{7} {Y_4}^{4} (\theta, \phi) \right\} \sin \pi (r - r_0),
\end{eqnarray}
for the cubic symmetric solution, 
where ${Y_\ell}^m (\theta, \phi)$ is the normalized spherical harmonic functions  
of degree $\ell$ and order $m$. 
Fig.~\ref{fig:yoshida01} shows the steady state convection pattern with the tetrahedral and cubic symmetries. 
We have performed benchmark tests 
with previously reported numerical mantle convection codes 
that employed various numerical schemes. 
In spite of the differences of the discretization methods, numerical techniques, 
and number of grid points among the codes, 
we found that the calculated values such as the Nusselt number 
obtained by our Yin-Yang mantle code agree well with previous calculations
within a few percent.

\subsection{Time-dependent convection}
\begin{figure}[ht]
\begin{center}
\noindent\includegraphics[width=0.8\linewidth]{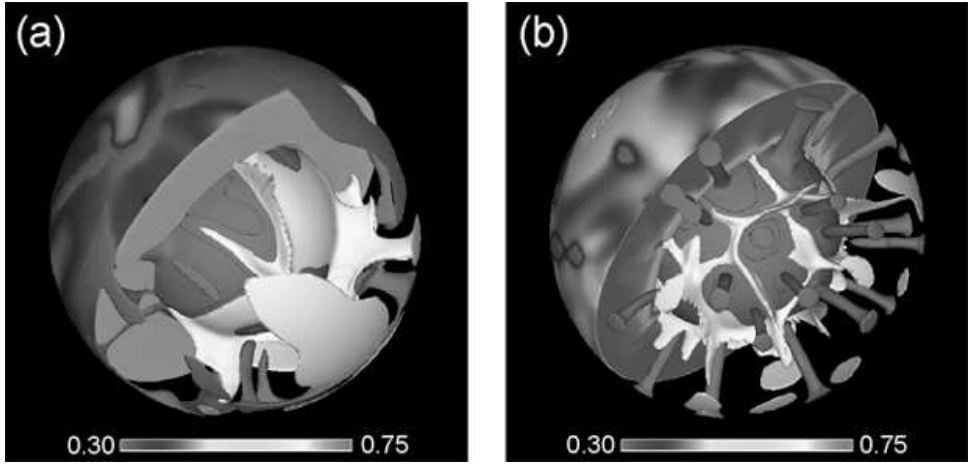}
\caption{
\label{fig:yoshida02}
The iso-surface of the temperature $T$ and the residual temperature $\delta T$ 
for the cases of (a) $H = 0$ and (b) $H = 20$. 
The Rayleigh number is $Ra = 10^7$. 
Iso-surfaces on the half spherical shell indicate the temperature (see color bars). 
Blue and Yellow iso-surfaces indicate $\delta T =  \pm 0.1$, respectively.  
Red spheres indicate the bottom of the mantle with fixed temperature.
}
\end{center}
\end{figure}

The Earth's mantle is obviously in a time-dependent convection 
under a very
high Rayleigh number ($Ra \geq 10^6$) and with internal heating ($H \leq 20$).  
When $Ra = 10^5$, the convection pattern becomes weakly time-dependent, 
and the geometrical symmetry is broken.  
Fig.~\ref{fig:yoshida02} shows the thermal structures of the mantle convection 
when $Ra = 10^7$ which is characteristic of the Earth's mantle. 
Without internal heating, 
the thermal structure is strongly time-dependent, driven by narrow, 
cylindrical upwelling (hot) plumes surrounding 
by a network of long downwelling (cold) sheets (Fig.~\ref{fig:yoshida02}a). 
This feature is in contrast with the convective feature at low Rayleigh number 
($Ra < 10^5$) where the convection is nearly steady state (Fig.~\ref{fig:yoshida01}). 
On the other hand, when the internal heating is taken into account ($H = 20$), 
the convective feature is dominated by 
the short-wavelength structure with numerous quasi-cylindrical downwellings  
spaced relatively close together. 
The downwellings are surrounded by a broad and diffuse upwelling of hotter fluid (Fig.~\ref{fig:yoshida02}b). 
We have found that internal heating has a 
strong influence on the scale and structure of the mantle convection,
especially on the shape of downwellings.

\begin{figure}[b]
\begin{center}
\noindent\includegraphics[width=0.8\linewidth]{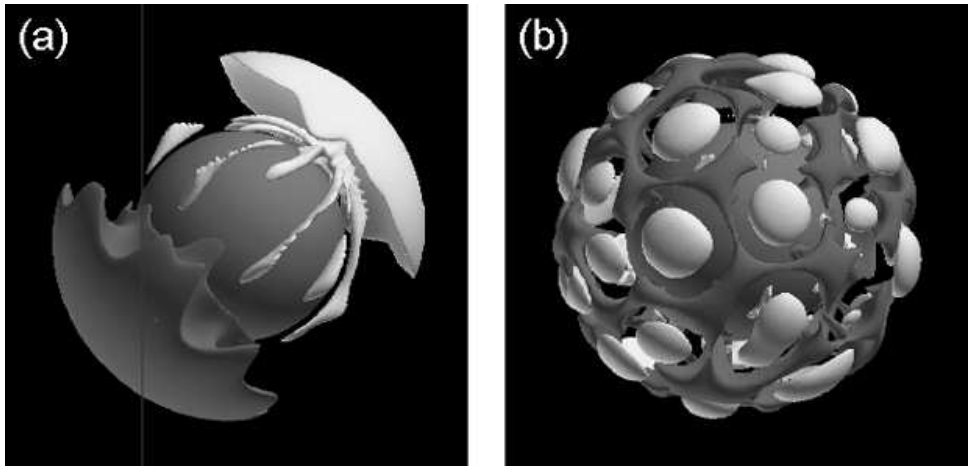}
\caption{
\label{fig:yoshida03}
The iso-surface of the residual temperature for the cases 
of (a) $\gamma_{mu} = 10^4$ $(E = 9.210)$ and (b) $\gamma_{\mu} = 10^6$ $(E = 13.816)$.  
Blue and Yellow iso-surfaces indicate 
(a) $\delta T = \pm 0.25$ and $\delta T = \pm 0.10$, respectively.  
Red spheres indicate the bottom of the mantle with fixed temperature. 
}
\end{center}
\end{figure}

The convection pattern is also drastically changed by taking the 
viscosity variation into account.
Fig.~\ref{fig:yoshida03} shows the thermal structures of the mantle convection 
with temperature-dependent viscosity at 
$Ra = 10^7$ and $H = 0$. 
When the temperature dependence of viscosity is rather moderate 
(the viscosity contrast across the convecting shell $\gamma_\mu$ is $10^3$--$10^4$), 
the convection has long-wavelength thermal structure 
with a mobile, stiff layer, or, ``sluggish-lid'' 
along the cold top surface of the mantle. 
When $Ra = 10^7$ and $\gamma_{\mu} = 10^4$, 
the convection pattern comes to be dominated by the degree-one pattern; 
the one cell structure that consists of a pair of cylindrical downwelling plume and 
cylindrical upwelling plume (Fig.~\ref{fig:yoshida03}a). 
On the other hand, the convective flow pattern that belongs to 
the ``stagnant-lid'' regime emerges when $\gamma_{\mu} \geq 10^5$. 
The stagnant-lid, which is an immobile, stiff layer, 
prevents the heat flux through the top boundary and 
leads to a small temperature difference in the mantle below the lid. 
Convection under the stagnant-lid is characterized 
by numerous, small-scale cylindrical plumes surroundings sheet-like downwelling (Fig.~\ref{fig:yoshida03}b). 
We have found that the variable viscosity with temperature dependence 
induces drastic effects on the mantle convection pattern.

\section{Application to geodynamo simulation}
The magnetic compass points to the north since the Earth is surrounded
by a dipolar magnetic field.
It is broadly accepted that the geomagnetic field is generated
by a self-excited electric current in the Earth's core,
The inner core is iron in solid state,
and the outer core is also iron but in liquid state due to the high temperature of the
planetary interior.
The electrical current is generated by
magnetohydrodynamic (MHD) dynamo
action---the energy conversion process from flow energy
into magnetic energy---of the liquid iron in the outer core.
In the last decade,
computer simulation has emerged as a central research method
for geodynamo study~\cite{kono_2002}.

In this section, we show the application
of the Yin-Yang grid to the geodynamo simulation
with a special emphasize on the code parallelization
and sustained performance achieved by the Earth Simulator.
We consider a spherical shell vessel bounded by two concentric
spheres.
The inner sphere of radius $r=r_i$ denotes the
inner core and the outer sphere of $r=r_o$ denotes the
core-mantle boundary.
An electrically conducting fluid is confined in this shell region.
Both the inner and outer spherical boundaries rotate with a constant
angular velocity ${\bf \Omega}$.
We use a rotating frame of reference with the same angular velocity.
There is a central gravity force in the direction of the center of
the spheres.
The temperatures of both the inner and outer spheres are fixed; hot
(inner) and cold (outer).
When the temperature difference is sufficiently large, a convection
motion starts when a random temperature perturbation is imposed at
the beginning of the calculation.
At the same time an infinitesimally small,
random ``seed'' of the magnetic field is given.

The system is described by the following normalized MHD equations:
\begin{equation} \label{eq:continuity}
     \frac{\partial \rho}{\partial t} = - \nabla \cdot \mathbf{f},
\end{equation}
\begin{equation} \label{eq:motion}
    \frac{\partial \mathbf{f}}{\partial t}
                   = -\nabla \cdot (\mathbf{v}\mathbf{f})
                     - \nabla p
       + {\bf j} \times {\bf B}
                     + \rho {\bf g}
                     + 2 \rho {\bf v}\times{\bf \Omega}
                     + \mu (   \nabla^2 {\bf v}
                     + \frac{1}{3} \nabla (\nabla \cdot {\bf v} ) ),
\end{equation}
\begin{equation} \label{eq:press}
         \frac{\partial p}{\partial t}
                = - \mathbf{v}\cdot \nabla p
                 - \gamma p \nabla \cdot {\bf v}
	              + (\gamma-1) K \nabla^2 T
		             +   (\gamma-1)\eta {\bf j}^2
			            +   (\gamma-1)\Phi,
\end{equation}
\begin{equation} \label{eq:induction}
   \frac{\partial {\bf A}}{\partial t} = \mathbf{v}\times\mathbf{B}+\eta\nabla^2\mathbf{A},
\end{equation}
with
\begin{eqnarray}
\nonumber
                &    p = \rho T, \hspace{3em}
                    {\bf B} =  \nabla \times {\bf A}, \hspace{2em}
                     {\bf j} =  \nabla \times {\bf B}, \hspace{2em}
                    {\bf g} = - g_0/r^2 {\bf \hat{r}}, \hspace{2em}\\
                &
                    \nabla \cdot \mathbf{B} = 0,\hspace{2em}
              \Phi =
                2 \mu \left( e_{ij} e_{ij}
	 - \frac{1}{3} (\nabla \cdot {\bf v})^2 \right), \hspace{2em}
       e_{ij} = \frac{1}{2} \left( \frac{\partial v_i}{\partial x_j}
       + \frac{\partial v_j}{\partial x_i}
          \right).
\end{eqnarray}
Here the mass density $\rho$, pressure $p$,
mass flux density $\mathbf{f}$,
magnetic field's vector potential $\mathbf{A}$ are
the basic variables in the simulation.
Other quantities;
magnetic field ${\bf B}$,
electric current density ${\bf j}$,
and electric field ${\bf E}$ are treated as subsidiary fields.
The ratio of the specific heat $\gamma$, viscosity $\mu$,
thermal conductivity $K$ and electrical resistivity $\eta$ are
assumed to be constant.
The vector ${\bf g}$ is the gravity acceleration and ${\bf \hat{r}}$ is the
radial unit vector; $g_0$ is a constant.
We normalize the quantities as follows:
The radius of the outer sphere $r_o = 1$; the temperature of the
outer sphere $T(1)$ = 1; and the mass density at the outer sphere
$\rho(1) = 1$.
The temperature on the inner and outer spheres are fixed.
The boundary condition for the velocity is rigid;
\begin{equation} \label{eq:1506}
  \mathbf{v} = 0, \ \ \ \ \hbox{at} \ r=r_i, 1.
\end{equation}
The boundary condition for the magnetic field is given by
\begin{equation} \label{eq:1508}
  B_\theta = B_\phi = 0,  \ \ \ \ \hbox{at} \ r=r_i, 1.
\end{equation}
We will consider the improvement of this rather artificial boundary
condition into more realistic one in the end of this section.
The spatial derivatives in the above equations are
discretized by the second-order central finite difference method
on the Yin-Yang grid.
The fourth-order Runge-Kutta method is used for
the temporal integration.
Initially,
both the convection energy
and the magnetic energy are negligibly small.
For geodynamo study, it is necessary
to follow the time development of the MHD system
until the thermal convection flow and the
dynamo-generated magnetic field are both sufficiently developed
and saturated.

We developed this Yin-Yang based geodynamo simulation code
for the Earth Simulator
by converting our previous geodynamo code, which was
based on the traditional latitude-longitude grid,
into the Yin-Yang grid.
We have found that
the code conversion from our previous latitude-longitude based code
into the new Yin-Yang based code is straightforward and rather easy.
Our experience with the rapid and easy conversion
from latitude-longitude code into Yin-Yang code
would be encouraging for others who have already developed
codes that are based on latitude-longitude
grids in the spherical coordinates,
and who are bothered by numerical problems and inefficiency
caused by the pole singularity.
We would like to suggest that they try the Yin-Yang grid.

\noindent
\begin{table}
\caption{\label{tab:1416}Specifications of the Earth Simulator.}
\begin{center}
\begin{tabular}{ll}
\br
Peak performance of arithmetic processor (AP) & 8 Gflops \\ 
Number of AP in a processor node (PN) & 8 \\ 
Total number of PN & 640 \\ 
Total number of AP & $8\hbox{ AP}\times 640\hbox{ PN}=5120$ \\ 
Shared memory size of PN & 16 GB \\ 
Total peak performance &
          $8 \hbox{ Gflops}\times 5120\hbox{ AP}=40 \hbox{Tflops}$ \\ 
Total main memory & 10 TB\\ 
Inter-node data transfer rate & 12.3 GB/s $\times$ 2 \\ 
\br
\end{tabular}
\end{center}
\end{table}

Since the Yin grid and Yang grid are identical,
dividing the whole computational domain into a Yin grid part
and a Yang grid part
is not only natural but also efficient for parallel processing.
In addition to this Yin-and-Yang division,
further domain decomposition within each grid
is applied to for the massively parallel computation
on the Earth Simulator.

The Earth Simulator, whose hardware specifications are
summarized in Table~\ref{tab:1416} has
three different levels of parallelization:
Vector processing in each arithmetic processor (AP);
shared-memory parallelization by 8 APs in each processor node (PN);
and distributed-memory parallelization by PNs.

In our Yin-Yang dynamo code,
we apply vectorization in the radial dimension
of the three-dimensional (3D) arrays for physical variables.
The radial grid size is 255 or 511,
which is just below the size (or doubled size) of
the vector register of
the Earth Simulator (256) to avoid
bank conflicts in the memory.
We use MPI both for the inter-node (distributed memory)
parallel processing
and for the intra-node (shared memory) parallel processing.
This approach is called ``flat-MPI'' parallelization.

As we mentioned above,
we first divide the
whole computational domain into two identical
parts that 
correspond to the Yin grid and Yang grid shown 
in Fig.~\ref{fig:yinyangBasic}(a).
(Therefore, the total number of processes is always even.)
For further parallelization within each component grid,
we applied the two-dimensional decomposition
in the horizontal space, colatitude $\theta$ and longitude $\phi$.
More details on the parallelization of this code is described in \cite{kageyama_2004b}.

The best performance of the Yin-Yang geodynamo code with the flat MPI
parallelization is $15.2$ Tflops.
This performance
is achieved by $4096$ processors (512 nodes) with the total grid
size of $511(\hbox{radial})
\times 514(\hbox{latitudinal})
\times 1538(\hbox{longitudinal})
\times 2(\hbox{Yin and Yang})$.
Since the theoretical peak performance of $4096$ processors
is $4096\times 8\hbox{ Gflops}=32.8 \hbox{Tflops}$,
we have achieved $46\%$ of peak performance in this case.
The average vector length is $251.6$,
and the vector operation ratio is $99\%$.
The high performance of the Yin-Yang dynamo code is
a direct consequence of
the simple and symmetric configuration design
of the Yin-Yang grid:
It makes it possible to minimize the
communication time ($10\%$) between the
processes in the horizontal directions,
and enables optimum vector processing (with $99\%$ of operation ratio)
in the radial direction in each process.

Before concluding this section, we briefly describe
our recent improvement of the Yin-Yang geodynamo code.
We have improved the boundary condition denoted by eq.~(\ref{eq:1508})
of the magnetic field into more realistic one, i.e., 
so called vacuum boundary condition.
In this boundary condition, the magnetic field generated by the
MHD dynamo in the outer core ($r\le 1$) is smoothly
connected to the magnetic field $\mathbf{B}_v$ of 
the outer region $r>1$ that is assumed to be an insulator;
\begin{equation} \label{eq:1553}
  \nabla \times \mathbf{B}_v = 0, \ \ \ \hbox{for} \ r>1.
\end{equation}
Therefore,  the $\mathbf{B}_v$ is written by a scalar function $\psi$,
\begin{equation} \label{eq:1607}
  \mathbf{B}_v = -\nabla \psi, \ \ \ \hbox{for} \ r>1,
\end{equation}
where, from $\nabla\cdot\mathbf{B}_v=0$, $\psi$ satisfies the potential equation
\begin{equation} \label{eq:1608}
  \nabla^2 \psi 
  = 
  \left[
  \frac{1}{r^2}\frac{\partial}{\partial r}\left(r^2 \frac{\partial}{\partial r}\right)
+ \frac{1}{r^2\sin\theta}\frac{\partial}{\partial\theta}
\left(\sin\theta\frac{\partial}{\partial\theta}\right)
+ \frac{1}{r^2 \sin^2\theta}\frac{\partial^2}{\partial \phi^2}
  \right] \psi
  =
  0, \ \ \ \hbox{for} \  r  \ge 1
\end{equation}

The boundary condition of $\psi$ at $r=1$ is given by
$-\nabla \psi(r=1)=B_r(r=1)$, where
$B_r(r=1)$ is determined from the dynamo region $(r\le 1)$.
Other component of the magnetic field at the surface $B_\theta(r=1)$ and
$B_\phi(r=1)$ are determined by the solution of eq.~(\ref{eq:1608}).
In order to solve this boundary value problem,
we first apply a coordinate transformation of $r$.
\begin{equation} \label{eq:1641}
  r \rightarrow \zeta = 1 / r.
\end{equation}
The equation~(\ref{eq:1608}) is converted into the
following form
\begin{equation} \label{eq:1646}
  \left[
  \zeta^2 \frac{\partial^2}{\partial \zeta^2}
+ \frac{1}{\sin\theta}\frac{\partial}{\partial\theta}
\left(\sin\theta\frac{\partial}{\partial\theta}\right)
+ \frac{1}{\sin^2\theta}\frac{\partial^2}{\partial \phi^2}
  \right] \psi
  =
  0, \ \ \ \hbox{for} \ 0 \le \zeta  \le 1
\end{equation}
The problem to solve eq.~(\ref{eq:1608}) \textit{outside} a unit sphere $r\ge1$
is now converted into
the problem to solve eq.~(\ref{eq:1646}) \textit{inside} a unit sphere $\zeta \le 1$.
The boundary condition of $\psi$ at the origin $\zeta=0$ is given by $\psi(\zeta=0)=0$
since $\psi(r=\infty)=0$.

\begin{figure}[ht]
\begin{center}
\noindent\includegraphics[width=\linewidth]{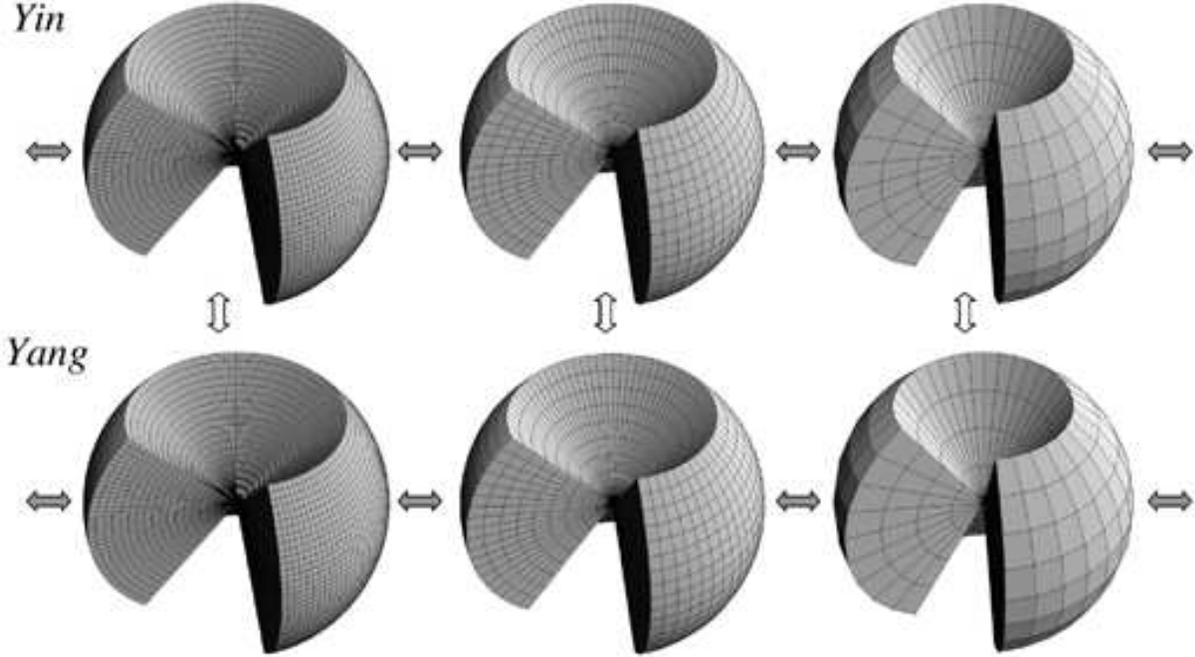}
\caption{
\label{fig:yymg}
The Yin-Yang multigrid method for the solution of the
vacuum magnetic field potential $\psi$.
The full approximation storage algorithm is used.
The horizontal boundary values of the Yin and Yang grids
for the overset are determined by the mutual interpolation
(white arrows)
at the every grid level in the V-cycle of the multigrid method (gray arrows).
}
\end{center}
\end{figure}

To solve eq.~(\ref{eq:1646}), we apply the multigrid method \cite{wesseling_2004},
which is practically the optimal way to solve this kind of boundary value
problem.
The base grid system is the Yin-Yang grid
defined in the full spherical region including the origin.
See Fig.~\ref{fig:yymg}.
We adopt the full approximation storage algorithm of the multigird method.
The Jacobi method is used as the smoother.
The V-cycle is repeated for a couple of times until we get the convergence.
The internal boundary condition of each component grid (Yin and Yang) 
are set by mutual bi-cubic interpolation at every grid level as 
indicated by white arrows in Fig.~\ref{fig:yymg}.
Although, the code is not parallelized yet,
its flat-MPI parallelization will be straightforward.
We have combined this non-parallelized Yin-Yang multigrid solver
of the vacuum potential $\psi$ with the
non-parallel version of the Yin-Yang geodynamo code.
We have found that the vacuum field condition has been successfully implemented
by this multigrid potential solver with almost the same computational cost (CPU time)
as with the MHD solver part.
This is a very promising result for further development.

\section{Summary}
We have developed a new spherical overset grid, ``Yin-Yang grid'', for 
geophysical simulations.
The Yin-Yang grid is constructed 
from a dissection of a sphere
into two identical and complemental pieces.
Among various possible overset grids over a sphere,
we believe that the Yin-Yang grid is the simplest and the most powerful
especially on massively parallel computers from the following reasons:
\begin{itemize}
\item 
It is an orthogonal system, since it is a part of the latitude-longitude grid.
\item
The grid spacing is quasi-uniform, since we picked up only 
the low latitude region of the latitude-longitude grid.
\item
The metric tensors are simple and analytically known,
since it is defined based on the spherical polar coordinates.
\item
Routines for the fluid (or MHD) solver 
can be recycled twice, since Yin and Yang are identical.
\item 
Routines for mutual interpolations of the overset grid borders can also be recycled twice,
since Yin and Yang are complemental.
\item
Parallelization is easy and efficient, since the domain decomposition is straightforward.
\end{itemize}

We have developed finite difference codes 
of the geodynamo simulation and the mantle convection simulation
on the Yin-Yang grid.
The Yin-Yang geodynamo code has achieved 15.2 Tflops with 4096
processors on the Earth Simulator.
This represents 46\% of the theoretical peak performance.
By the Yin-Yang mantle code, we can carry out
realistic mantle convection simulations under
the Rayleigh number of $10^7$, including
strongly temperature-dependent viscosity whose 
contrast reaches upto $10^6$.

In the Earth Simulator Center, 
the Yin-Yang grid is also applied to advanced 
general circulation modes of the atmosphere and ocean \cite{takahashi_2004, komine_2004,takahashi_2004b}.

\ack
The authors would like to thank Prof.~Tetsuya Sato, the director-genenal of the
Earth Simulator Center, for instructive discussions 
and Dr.~Masanori Kameyama for
useful comments on the application of the multigrid method to the Yin-Yang grid.
All the simulations were performed on the Earth Simulator.

\section*{References}


\end{document}